# Combinative Isobaric Resonances CIRs by MSS and from previous precise data.

A. Gafarov[1*]

[1]Nuclear Reactions Laboratory, INP of Science Academy of Republic Uzbekistan
[*]email: anatxagor@gmail.com

The elastic scattering $^{12}$C(p,p$_o$) Excitation Function (EF) was measured in the energy range $E_p$ =16 ÷ 19.5 MeV with resolution ~10 keV by means of innovative approach - **MSS** - the *Method of Spectra Superposition* at the 14-angle Magnetic Spectrograph (Apelsin) at beam with an energy spread of ~200 keV from U-150 cyclotron of INP Ulugbek (Tashkent) of Science Academy of Uzbekistan [1-9]. The obtained EF has a reach structure of anomalies in precise agreement with thresholds & levels data [2,10-12]. Measurements were done on the $^{12}$C self-supporting target with thickness 13 mg/cm$^2$ and an area of 1 cm$^2$ in the center of Apelsin. The 20-step Energy Moderator controlled the proton energy, providing a 3.5 MeV wide $E_p$ interval with no readjustment of cyclotron and all the ionic-optics on the 41-meter-long beam-pipe. Energy resolution of Apelsin for protons is better than 5 keV. The spectrograph magnetic field was stabilized to 3ppm by NMR-monitor system [1,9]. The EF of $^{12}$C (p,p$_o$) scattering has many resonances that precisely correspond to g.s. and levels of well-known product-nuclei. But 80% of anomalies were not decoded.
A new concept of Combinative Isobaric Resonance (CIRs) proposed to identify these anomalies. The proposed CIRs concept worked very promising (Fig. 4). To prove CIRs validity, precise excitation function of $^{12}$C(p,p$_o$), $^{12}$C(d,d$_o$), and $^7$Li(p,p$_o$), were analyzed in search for the Combinative Isobaric Resonances. This study perfectly proves existence of CIRs.

## 1. Introduction

Many low-energy precise excitation functions measured at Van-de-Graaf and tandem accelerators have displayed multiple anomalies with a fraction of ones never clearly identified yet.

It appeared again when by the MSS first precise excitation function for (**p,$^{12}$C**)-elastic scattering got measured at magnetic spectrograph "Apelsin" in the energy range 16÷19.5 MeV with ΔE ~10 keV [1-9].

It is obvious when looking on the Fig. 1, where only ~20% of structures are well-known peaks which are precisely located under the vertical lines of the corresponding g.s. and levels of excited states of product-nuclei from reliable data sources [10-13].

## 2. The familiar products of the ($p + {}^{12}C$ )

The Reading 0 is the key expression for calculation of the verticals on Fig. 1 for traces of population of all known product-nuclei in the ground or excited states:

$$E_{Pcm} = E^*_{LNP} + E^*_{RC} + \Delta M_{OS} - \Delta M_{PS}, \qquad (1)$$

$$\Delta M_{OS} = \Delta M_{NP} + \Delta m_{RC} ,$$

$$\Delta M_{PS} = \Delta M_{TN} + \Delta m_{BP} ,$$

where $E_{Pcm}$ -is the projectile's energy in CM-frame, $E^*_{LNP}$ -is the excitation energy of the product-nucleus (level energy), $E^*_{RC}$ -is the excitation energy of the residual cluster, $\Delta M_{OS}$-is the mass-excess of the output system, $\Delta M_{PN}$ -is the mass-excess of the product-nucleus, $\Delta m_{RC}$ -is the mass-excess of the residual cluster, $\Delta M_{PS}$ -is the mass-excess of primary system, $\Delta M_{TN}$ -is the mass-excess of the target-nucleus, $\Delta m_{BP}$ -is the mass-excess of the beam-particle.

All anomalies in the EF are proportional to $\sigma$ (rates) of population of the nearest energy-accessible "cells" in the phase volume of the ($p + {}^{12}C$ ) interaction, even no other than elastic product-$p$ is detected.

It works on the schematic:

**Primary system → Populated system → decay into primary system**

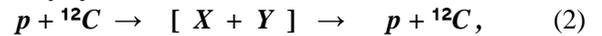
$$p + {}^{12}C \rightarrow [\ X + Y\ ] \rightarrow p + {}^{12}C, \qquad (2)$$

First candidates, to appear as resonances in the EF, are the levels of $^{12}C^*$ that protons with 16 MeV ÷19.5 MeV can populate:

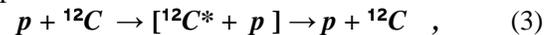
$$p + {}^{12}C \rightarrow [{}^{12}C^* + p\ ] \rightarrow p + {}^{12}C \quad, \qquad (3)$$

The $^{12}C^*$ diagram of energy levels is right over the EF on Fig. 1, and the corresponding peaks are 15.110 MeV, very bright 16.1067 MeV, 16.58 MeV, 17.23 MeV, and 17.76 MeV. The triangle bars represent known width of each level and error bars are right below, they cross vertical lines.

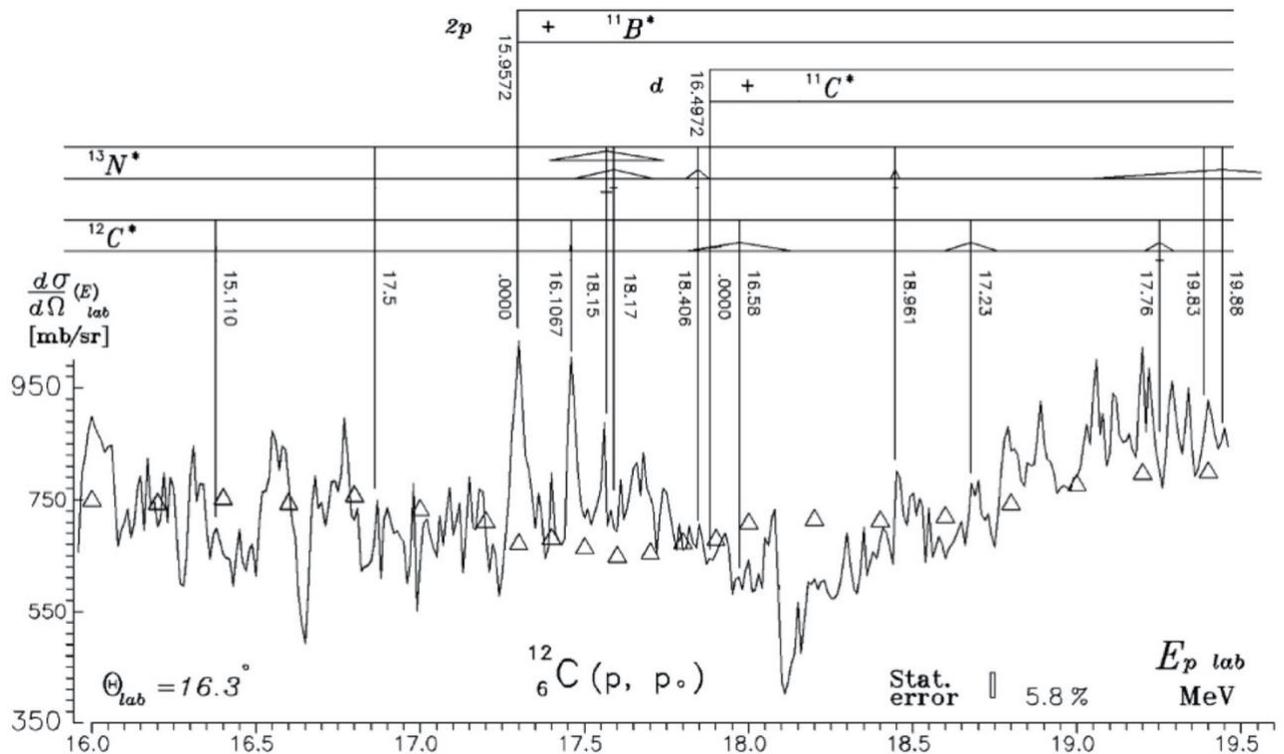

Fig. 1. Excitation function of $^{12}C(p, p_o)$ measured in the range 16 ÷ 19.5 MeV by MSS-approach at magnetic spectrograph "Apelsin" and U-150 cyclotron beam with $E_p$ =19.5 MeV with energy-spread ~200 keV.

Δ - from [13] W. W. Daehnick and R. Sherr. Phys. Rev. 133, B934

The brightest peak in the EF middle corresponds to:

$$p + {}^{12}C \rightarrow [{}^{11}B^* + 2p] \rightarrow {}^{12}C + p, \qquad (4)$$

It precisely seats under the vertical $^{11}B$ g.s. line, which is 15.9572 MeV in the rest frame.
Population of $^{13}N^*$ brings some bright resonances:

$$p + {}^{12}C \rightarrow [{}^{13}N^*] \rightarrow {}^{12}C + p, \qquad (5)$$

Also $^{11}C$ g.s. ( + $d$ ) is right in the middle of. Fig.1.

Other products have the g.s. and levels outside the proton energy range of 16 MeV ÷ 19.5 MeV. Here below they are:

$$p + {}^{12}C \rightarrow [\,p + {}^{4}He + {}^{4}He + {}^{4}He\,] \rightarrow p + {}^{12}C, \quad (6)$$
$$p + {}^{12}C \rightarrow [\,p + {}^{8}Be^* + {}^{4}He\,] \rightarrow p + {}^{12}C, \quad (7)$$
$$p + {}^{12}C \rightarrow [\,{}^{9}Be^* + {}^{4}He^*\,] \rightarrow p + {}^{12}C, \quad (8)$$
$$p + {}^{12}C \rightarrow [\,{}^{5}Li + {}^{4}He + {}^{4}He\,] \rightarrow p + {}^{12}C, \quad (9)$$
$$p + {}^{12}C \rightarrow [\,{}^{8}Be^* + {}^{5}Li\,] \rightarrow p + {}^{12}C, \quad (10)$$

**The Fig. 1 looks kind of strange. The unrecognized peaks (80%) are bigger than 3σ, but what they are?** The statistical error is less than 6%. They must be something that matters!

## 2. The Combinative Isobaric Resonances (CIRs)-concept.
(The source of additional resonances)

According to the CIR-concept, the projectile acts by populating the **nearest energy-accessible state in the A=13 phase volume**. A13 accommodates the incoming energy **by populating/creating** of an **initiated** nuclear combination, which soon will decay into the primary system [$^{12}C$ g.s. + $p$] with the $E_p$ of elastic $p$ -peak.
The populated combination comprises an "initiated" nucleus + some residual cluster, and all energy-accessible isobaric combination can play "this game".
   Let's, just for fun, check all **combinations of nuclei + residuals** which give **A=13**, temporarily ignoring the charge conservation…

One can use the same expression (1) in *Writing I* form to calculate positions of levels and ground states.
*Writing I*:

$$E_{Pcm} = E^*_{LIN} + E^*_{RC} + \Delta M_{IS} - \Delta M_{PS}, \qquad (11)$$

$$\Delta M_{IS} = \Delta M_{IN} + \Delta m_{RC},$$

$$\Delta M_{PS} = \Delta M_{TN} + \Delta m_{BP},$$

where $E_{Pcm}$ -is the projectile's energy in CM-frame, $E^*_{LIN}$ -is the excitation energy of the initiated nucleus (level energy), $E^*_{RC}$ -is the excitation energy of the residual cluster, $\Delta M_{IS}$ -is the initiated system mass-excess, $\Delta M_{IN}$ -is the mass-excess of the initiated nucleus, $\Delta m_{RC}$ -is the mass-excess of the residual cluster, $\Delta M_{PS}$ -is the mass-excess of primary system, $\Delta M_{TN}$ -is the mass-excess of the target-nucleus, $\Delta m_{BP}$ -is the mass-excess of the beam-particle.

**Just for fun, (3) also is going to split into isobaric combinations below:**

**primary system, initiated nucleus & cluster    primary system**

$$p + {}^{12}C \rightarrow [\ {}^{12}Li^* + p\ ] \rightarrow p + {}^{12}C, \quad (12)$$
$$p + {}^{12}C \rightarrow [\ {}^{12}Be^* + p\ ] \rightarrow p + {}^{12}C, \quad (13)$$
$$p + {}^{12}C \rightarrow [\ {}^{12}B^* + p\ ] \rightarrow p + {}^{12}C, \quad (14)$$
$$p + {}^{12}C \rightarrow [\ {}^{12}C^* + p\ ] \rightarrow p + {}^{12}C, \quad (3)$$
$$p + {}^{12}C \rightarrow [\ {}^{12}N^* + p\ ] \rightarrow p + {}^{12}C, \quad (15)$$
$$p + {}^{12}C \rightarrow [\ {}^{12}O^* + p\ ] \rightarrow p + {}^{12}C, \quad (16)$$
$$p + {}^{12}C \rightarrow [\ {}^{12}F^* + p\ ] \rightarrow p + {}^{12}C, \quad (17)$$
$$p + {}^{12}C \rightarrow [\ {}^{12}Ne^* + p\ ] \rightarrow p + {}^{12}C, \quad (18)$$

Set (4) is going to split into isobaric combinations:

$$p + {}^{12}C \rightarrow [\ {}^{11}Li^* + 2p] \rightarrow {}^{12}C^* + p, \quad (19)$$
$$p + {}^{12}C \rightarrow [{}^{11}Be^* + 2p] \rightarrow {}^{12}C^* + p, \quad (20)$$
$$p + {}^{12}C \rightarrow [\ {}^{11}B^* + 2p] \rightarrow {}^{12}C^* + p, \quad (4)$$
$$p + {}^{12}C \rightarrow [\ {}^{11}C^* + 2p] \rightarrow {}^{12}C^* + p, \quad (21)$$
$$p + {}^{12}C \rightarrow [\ {}^{11}N^* + 2p] \rightarrow {}^{12}C^* + p, \quad (22)$$
$$p + {}^{12}C \rightarrow [\ {}^{11}O^* + 2p] \rightarrow {}^{12}C^* + p, \quad (23)$$
$$p + {}^{12}C \rightarrow [\ {}^{11}F^* + 2p] \rightarrow {}^{12}C^* + p, \quad (24)$$
$$p + {}^{12}C \rightarrow [{}^{11}Ne^* + 2p] \rightarrow {}^{12}C^* + p, \quad (25)$$

Set (5) is going to split into isobaric combinations:

$$p + {}^{12}C \rightarrow [\ {}^{13}Li^*] \rightarrow {}^{12}C^* + p, \quad (26)$$
$$p + {}^{12}C \rightarrow [\ {}^{13}Be^*] \rightarrow {}^{12}C^* + p, \quad (27)$$
$$p + {}^{12}C \rightarrow [\ {}^{13}B^*\ ] \rightarrow {}^{12}C^* + p, \quad (28)$$
$$p + {}^{12}C \rightarrow [\ {}^{13}C^*\ ] \rightarrow {}^{12}C^* + p, \quad (5)$$
$$p + {}^{12}C \rightarrow [\ {}^{13}N^*\ ] \rightarrow {}^{12}C^* + p, \quad (29)$$
$$p + {}^{12}C \rightarrow [\ {}^{13}O^*\ ] \rightarrow {}^{12}C^* + p, \quad (30)$$
$$p + {}^{12}C \rightarrow [\ {}^{13}F^*\ ] \rightarrow {}^{12}C^* + p, \quad (31)$$
$$p + {}^{12}C \rightarrow [\ {}^{13}Ne^*] \rightarrow {}^{12}C^* + p, \quad (32)$$

Set (6) is going to split into isobaric combinations:

$$p + {}^{12}C \rightarrow [\,p + {}^{4}He + {}^{4}He + {}^{4}He\,] \rightarrow p + {}^{12}C, \quad (6)$$
$$p + {}^{12}C \rightarrow [\,p + {}^{4}Li + {}^{4}He + {}^{4}He\,] \rightarrow p + {}^{12}C, \quad (33)$$
$$p + {}^{12}C \rightarrow [\,p + {}^{4}Li + {}^{4}Li + {}^{4}He\,] \rightarrow p + {}^{12}C, \quad (34)$$
$$p + {}^{12}C \rightarrow [\,p + {}^{4}Li + {}^{4}Li + {}^{4}Li\,] \rightarrow p + {}^{12}C, \quad (35)$$

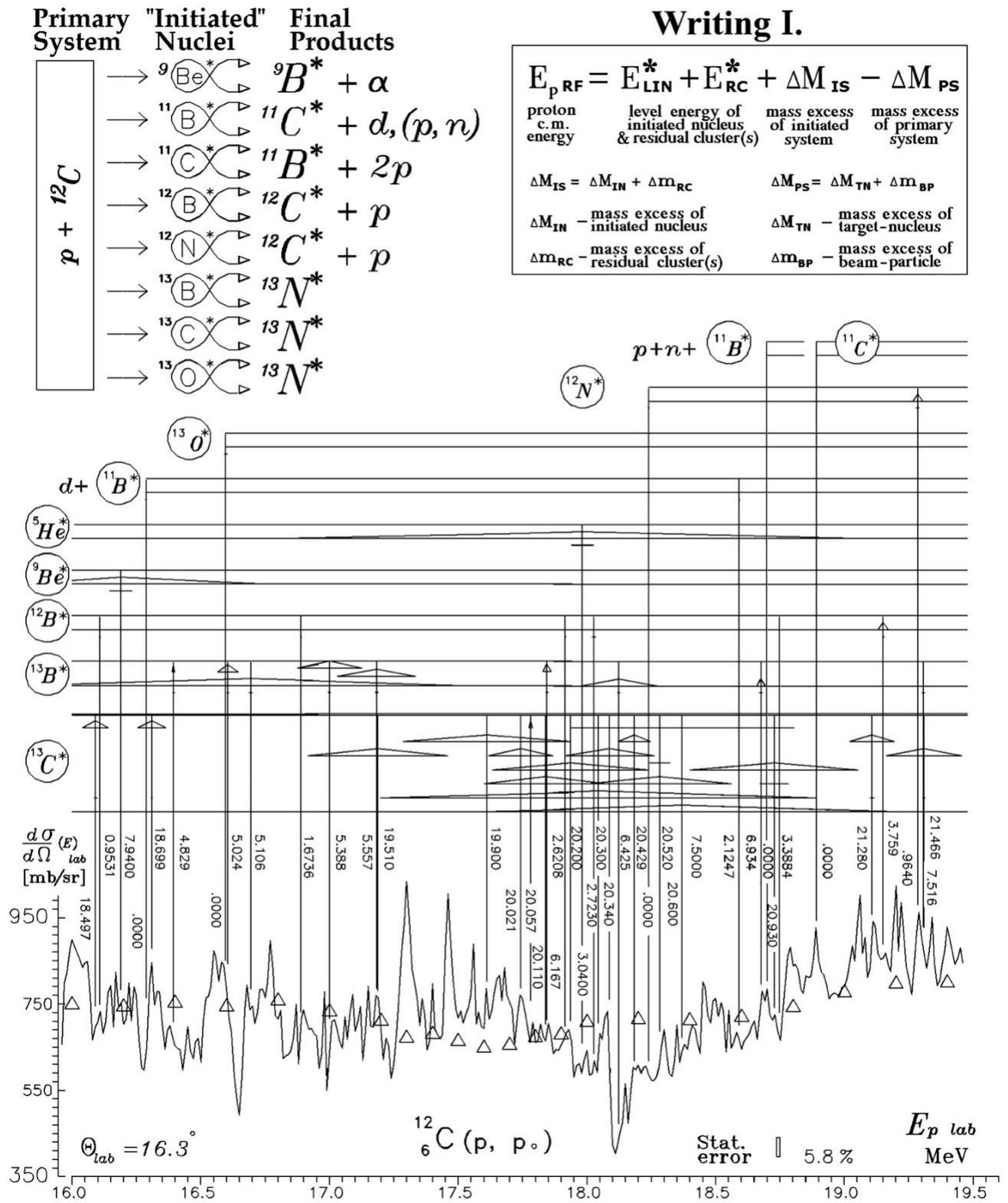

Fig. 2. Excitation function of $^{12}C(p, p_o)$ measured in the range 16 ÷ 19.5 MeV by MSS-approach at magnetic spectrograph "Apelsin" and U-150 cyclotron beam with $E_p$ =19.5 MeV with energy-spread ~200 keV.

Set (7) is going to split into isobaric combinations:

$p + {}^{12}C \rightarrow [\, p \,+\, {}^{8}Be^* +\, {}^{4}He\,] \rightarrow p + {}^{12}C,$ ( 7)
$p + {}^{12}C \rightarrow [\, p \,+\, {}^{8}Be^* +\, {}^{4}Li\,] \rightarrow p + {}^{12}C,$ (36)
$p + {}^{12}C \rightarrow [\, p \,+\, {}^{8}Li^* +\, {}^{4}He\,] \rightarrow p + {}^{12}C,$ (37)
$p + {}^{12}C \rightarrow [\, p \,+\, {}^{8}Li^* +\, {}^{4}Li\,] \rightarrow p + {}^{12}C,$ (38)
$p + {}^{12}C \rightarrow [\, p \,+\, {}^{8}B^* +\, {}^{4}He\,] \rightarrow p + {}^{12}C,$ (39)
$p + {}^{12}C \rightarrow [\, p \,+\, {}^{8}B^* +\, {}^{4}Li\,] \rightarrow p + {}^{12}C,$ (40)
$p + {}^{12}C \rightarrow [\, p \,+\, {}^{8}C^* +\, {}^{4}He\,] \rightarrow p + {}^{12}C,$ (41)
$p + {}^{12}C \rightarrow [\, p \,+\, {}^{8}C^* +\, {}^{4}Li\,] \rightarrow p + {}^{12}C,$ (42)

Set (8) is going to split into isobaric combinations:

$p + {}^{12}C \rightarrow [\, {}^{9}Be^* +\, {}^{4}He^*\,] \rightarrow p + {}^{12}C,$ ( 8)
$p + {}^{12}C \rightarrow [\, {}^{9}Be^* +\, {}^{4}Li^*\,] \rightarrow p + {}^{12}C,$ (43)
$p + {}^{12}C \rightarrow [\, {}^{9}Li^* +\, {}^{4}He^*\,] \rightarrow p + {}^{12}C,$ (44)
$p + {}^{12}C \rightarrow [\, {}^{9}Li^* +\, {}^{4}Li^*\,] \rightarrow p + {}^{12}C,$ (45)
$p + {}^{12}C \rightarrow [\, {}^{9}B^* +\, {}^{4}He^*\,] \rightarrow p + {}^{12}C,$ (46)
$p + {}^{12}C \rightarrow [\, {}^{9}B^* +\, {}^{4}Li^*\,] \rightarrow p + {}^{12}C,$ (47)
$p + {}^{12}C \rightarrow [\, {}^{9}C^* +\, {}^{4}He^*\,] \rightarrow p + {}^{12}C,$ (48)
$p + {}^{12}C \rightarrow [\, {}^{9}C^* +\, {}^{4}Li^*\,] \rightarrow p + {}^{12}C,$ (49)
$p + {}^{12}C \rightarrow [\, {}^{9}N^* +\, {}^{4}He^*\,] \rightarrow p + {}^{12}C,$ (50)
$p + {}^{12}C \rightarrow [\, {}^{9}N^* +\, {}^{4}Li^*\,] \rightarrow p + {}^{12}C,$ (51)

Set (9) is going to split into isobaric combinations:

$p + {}^{12}C \rightarrow [\, {}^{5}Li + {}^{4}He + {}^{4}He\,] \rightarrow p + {}^{12}C,$ ( 9)
$p + {}^{12}C \rightarrow [\, {}^{5}Li + {}^{4}Li + {}^{4}He\,] \rightarrow p + {}^{12}C,$ (52)
$p + {}^{12}C \rightarrow [\, {}^{5}Li + {}^{4}Li + {}^{4}Li\,] \rightarrow p + {}^{12}C,$ (53)
$p + {}^{12}C \rightarrow [\, {}^{5}Be + {}^{4}He + {}^{4}He\,] \rightarrow p + {}^{12}C,$ (54)
$p + {}^{12}C \rightarrow [\, {}^{5}Be + {}^{4}Li + {}^{4}He\,] \rightarrow p + {}^{12}C,$ (55)
$p + {}^{12}C \rightarrow [\, {}^{5}Be + {}^{4}Li + {}^{4}Li\,] \rightarrow p + {}^{12}C,$ (56)

Set (10) is going to split into isobaric combinations:

$p + {}^{12}C \rightarrow [\, {}^{8}Be^* + {}^{5}Li\,] \rightarrow p + {}^{12}C,$ (10)
$p + {}^{12}C \rightarrow [\, {}^{8}Be^* + {}^{5}He\,] \rightarrow p + {}^{12}C,$ (57)
$p + {}^{12}C \rightarrow [\, {}^{8}Li^* + {}^{5}He\,] \rightarrow p + {}^{12}C,$ (58)
$p + {}^{12}C \rightarrow [\, {}^{8}Li^* + {}^{5}Li\,] \rightarrow p + {}^{12}C,$ (59)
$p + {}^{12}C \rightarrow [\, {}^{8}B^* + {}^{5}Li\,] \rightarrow p + {}^{12}C,$ (60)
$p + {}^{12}C \rightarrow [\, {}^{8}B^* + {}^{5}He\,] \rightarrow p + {}^{12}C,$ (61)
$p + {}^{12}C \rightarrow [\, {}^{8}C^* + {}^{5}Li\,] \rightarrow p + {}^{12}C,$ (62)
$p + {}^{12}C \rightarrow [\, {}^{8}C^* + {}^{5}He\,] \rightarrow p + {}^{12}C,$ (63)
$p + {}^{12}C \rightarrow [\, {}^{8}N^* + {}^{5}Li\,] \rightarrow p + {}^{12}C,$ (64)
$p + {}^{12}C \rightarrow [\, {}^{8}N^* + {}^{5}He\,] \rightarrow p + {}^{12}C,$ (65)

**So, after calculation by *Writing I* of vertical pointers for all the combinations (3-65) no mismatches were observed – only good agreements, see Fig.2 !!!   Very promising.**

The Reading 0 (1) and *Writing I* (11) together cover about 60% of all peaks (Fig.1 and Fig.2).

They both do not look obviously including the weak/electromagnetic forces involved (only strong forces?). How to fix it?

An additional expression was proposed with obvious terms of the weak and electromagnetic contributions – the *Writing II* :

$$E_{Pcm} = E^*_{LIN} + E^*_{RC} + \Delta M_{FS} - \Delta M_{PS} + E_{NUC}, \quad (66)$$

$$\Delta M_{FS} = \Delta M_{FN} + \Delta m_{RC},$$
$$\Delta M_{PS} = \Delta M_{TN} + \Delta m_{BP},$$

$$E_{NUC} = \Delta E_{IN} - \Delta E_{FN},$$

$I$ and $F$ - Initiated and Final systems respectively,

$$\Delta E_{IN} = \Delta M_{IN} - Z_I \Delta m_p - N_I \Delta m_n - 0.6 Z_I (Z_I - 1) \sqrt[3]{A_I},$$
$$\Delta E_{FN} = \Delta M_{FN} - Z_F \Delta m_p - N_F \Delta m_n - 0.6 Z_F (Z_F - 1) \sqrt[3]{A_F},$$

where $E_{Pcm}$ -is the projectile's energy in CM-frame, $E^*_{LIN}$ -is the excitation energy of the initiated nucleus (level energy), $E^*_{RC}$ -is the excitation energy of the residual cluster, $\Delta M_{IS}$-is the initiated system mass-excess, $\Delta M_{IN}$ -is the mass-excess of the initiated nucleus, $\Delta m_{RC}$ -is the mass-excess of the residual cluster, $\Delta M_{PS}$ -is the mass-excess of primary system, $\Delta M_{TN}$ -is the mass-excess of the target-nucleus, $\Delta m_{BP}$ -is the mass-excess of the beam-particle, $I$ and $F$ -Initiated and Final nuclear systems, $Z$ and $N$ protons and neutrons and $\Delta m_p$, $\Delta m_n - p, n$ mass-defects.

The (66) meaning is that population happens with weak and electromagnetic interactions involved too.

Application of (66) to combinations (3-65) resulted in Fig.3, where agreement is impressing once again.

Let's put all together Reading 0, *Writing I* & *Writing II* on a diagram to have a general view (Fig 4).

As one can see, the general view now looks very promising, **but how this could be**?

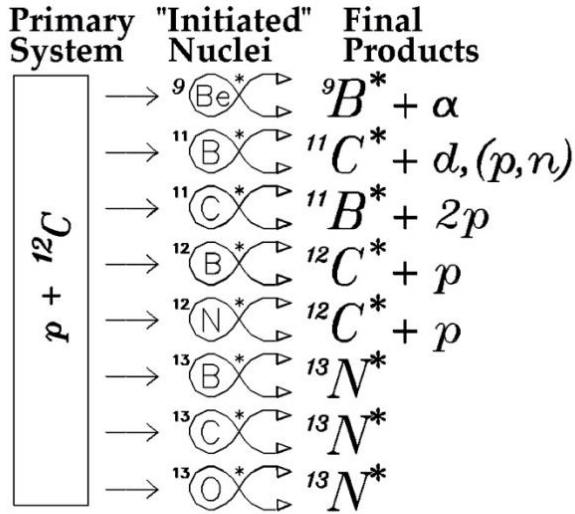

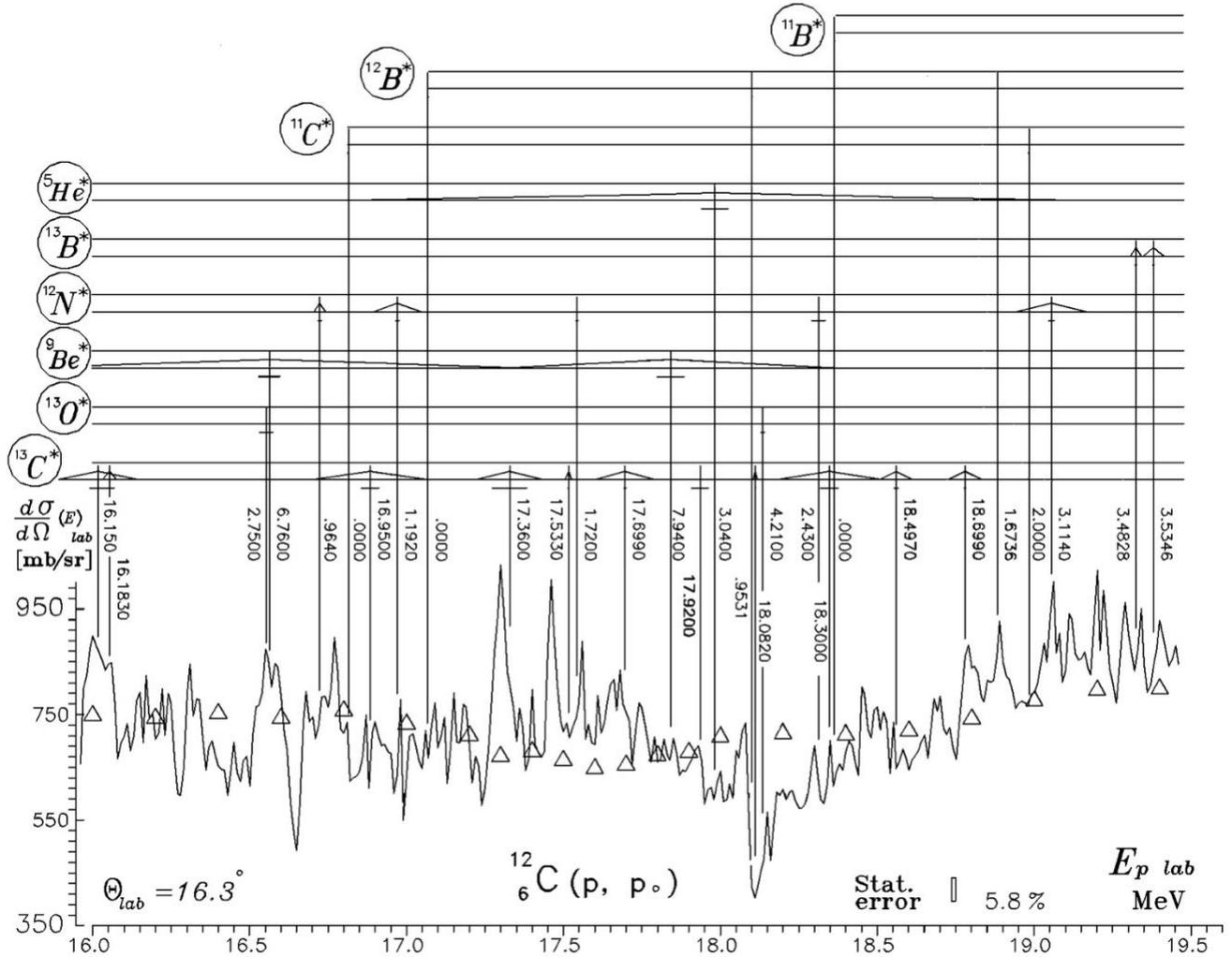

Fig. 3. Excitation function of $^{12}C(p, p_o)$ measured in the range $16 \div 19.5$ MeV by MSS-approach at magnetic spectrograph "Apelsin" and U-150 cyclotron beam with $E_p$ =19.5 MeV with energy-spread ~200 keV.
**CIRs via weak&EM forces -Writing II.** Δ - [13] W. W. Daehnick and R. Sherr. Phys. Rev. 133, B934

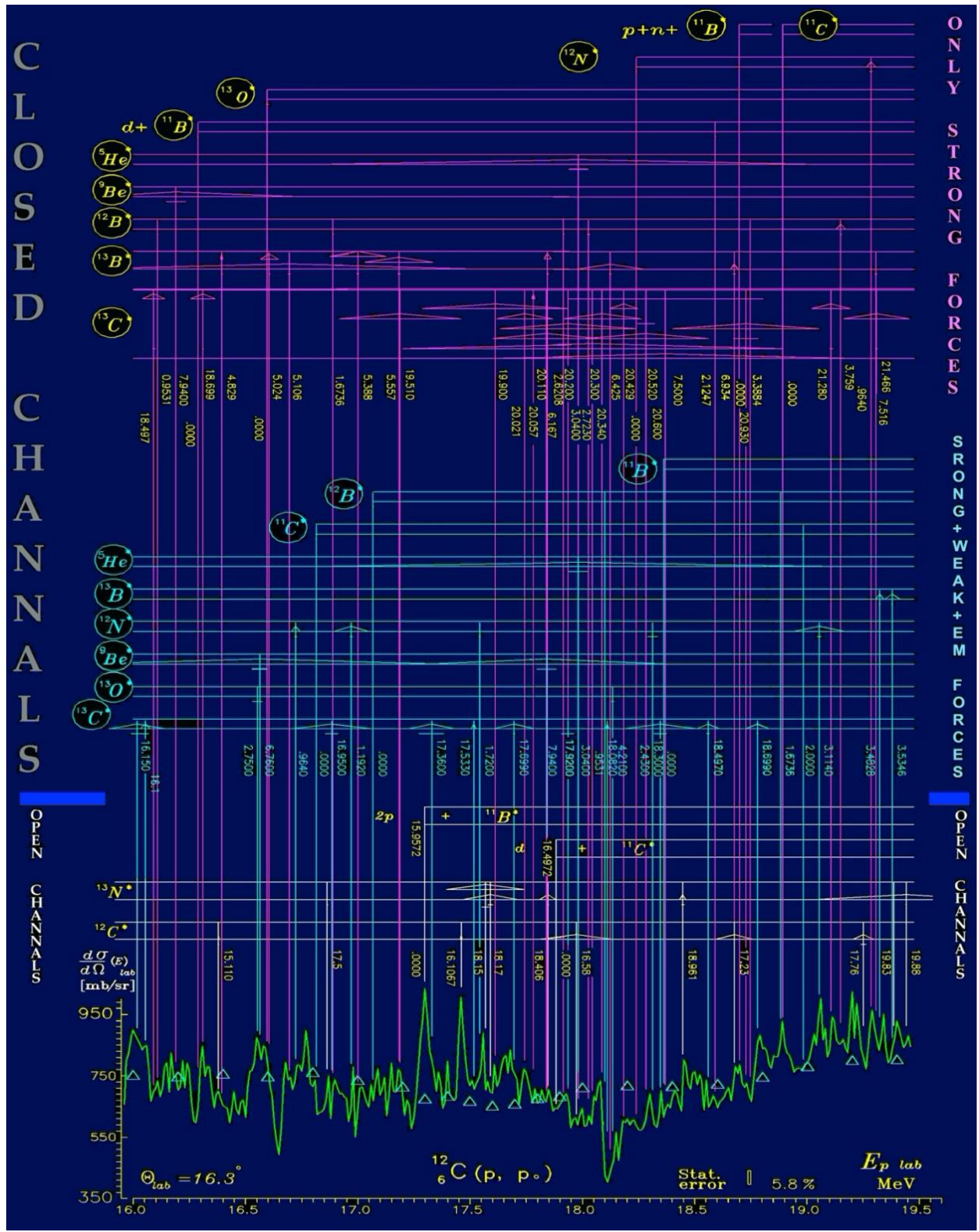

Fig. 4. The EF of $^{12}C(p, p_o)$ in the range $16 \div 19.5$ MeV by the MSS at magnetic spectrograph "Apelsin".
Possible combination isobaric resonances in A13 population according to Reading 0, Writing I and Writing II
Δ - from [13]. W. W. Daehnick and R. Sherr. Phys. Rev. 133, B934

Really, the electric charge conservation is a strong law.

**What if population in the A13 phase volume happens with synchronic production of "charge balancing" particles deeply bound in nuclear medium???**

What's known – the nuclear medium self-supports via the exchange-currents which wrap around every nucleon in the nucleus - it's a 3D-nuclear **binding net** -a honeycomb of strong field that, being disturbed, makes $\pi$ / W -sounds, propagating and then quenching in nucleus [14,15].

Production of deeply bound $\pi$ is in study now [16-20]. But, it's for heavy nuclei. How about light A13???

Well, even a neutron alone polarize vacuum so much that, by a **critical vacuum fluctuation**, neutron derives from vacuum a W-boson which turns **n** to decay in a **p**.

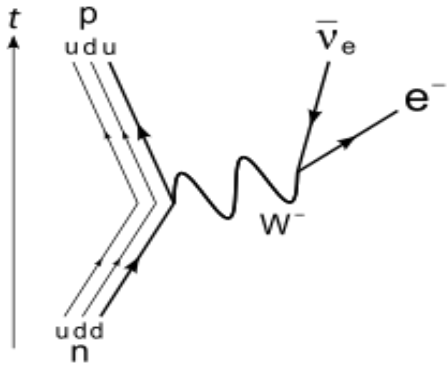

Fig.5.

Mass of **n** is ~1 GeV, but W itself is heavier than 80 GeV. Even a neutron alone can derive huge things from vacuum.

Why a **12C** nucleus **+ p** -strike, cannot???  **It can!**
"Overall" strong field of **12C** is ~12 GeV. Just disturb it and it will respond with a bunch of deeply bound particles ($\pi$ or W sounds [14], sometimes internally branching into $\mu$-mesons and/or leptons. In the nuclear medium the branching ratios differ from the known ones).

Let's make a special Feynman diagram for population of [ **13N g.s /*+$\pi^o$**] initiated system according to *Reading 0*:

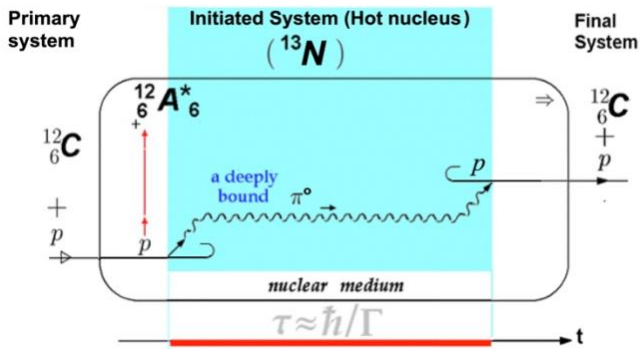

Fig. 6. Population of [**13N g.s** or* +$\pi^o$] initiated system.

It's has to be noticed, there are no **13N** initiation attempts according to *Writing I* & *Writing II*.

## 3. The "charge rising" CIRs

According to the *Writing I*
Population of a **13O** initiated nucleus (in **g.s.** or excited) with a deeply bound $\pi^-$ inside is below:

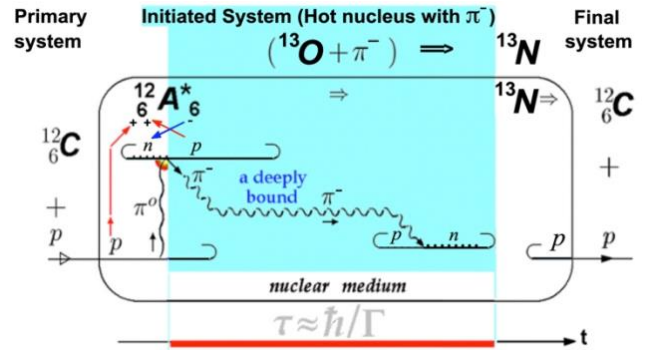

Fig. 7. Population of [**13O g.s./*+$\pi^-$**] initiated system.

According to the *Writing II*
Population of a **13O** initiated nucleus (**g.s.** / *) with deeply bound

$$W^- \to [\mu^- + \bar{\nu}_\mu] \to e^- + \bar{\nu}_e , \quad \text{or} \quad (67)$$

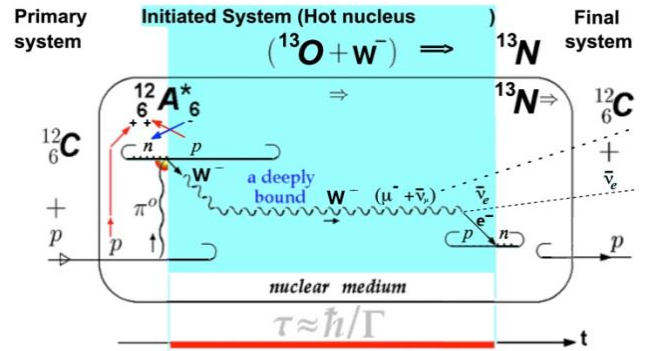

Fig. 8.
or with $\quad \pi^- \to [\mu^- + \bar{\nu}_\mu] \to e^- + \bar{\nu}_e$ : (68)

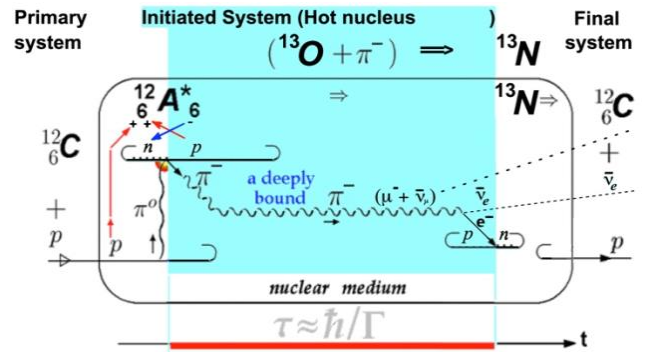

Fig. 9.

To see the other isobars $^{13}F$ and $^{13}Ne$, the **p**-projectile has to have enough energy to stimulate two (or three) synchronic transition of **n→p** (the return **p→n** transitions with 2 (or 3) of $\pi^-$, $\mu^-$ or $e^-$, will follow automatically). This is the "**charge rising**" branching.

According to *Writing I*, the charge rising population of the $^{13}F$ states could look like this below:

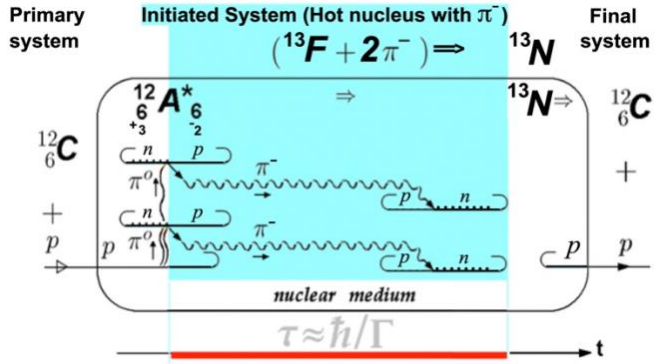

Fig.10.

According to *Writing II*, the charge rising population of the $^{13}F$ could look like diagram below:

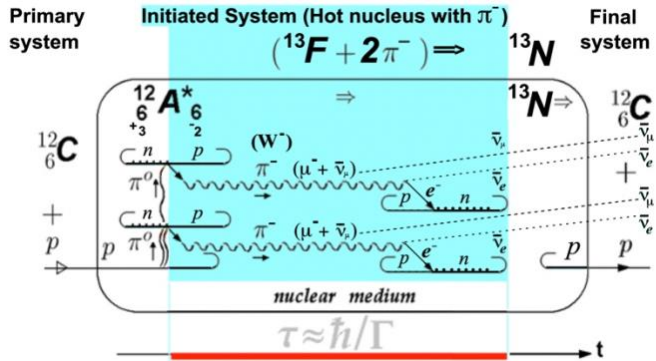

Fig.11.

The upper diagram could have $W^-$ instead of $\pi^-$.

Presence of neutrinos can cause some widening of the resonance because of recoil, similar to the $\beta$-decay, but less ($\pi$ and $\mu$ are much heavier than $\beta$-particle).

## 4. The "charge lowering" CIRs

To see population on the "**charge lowering**" branching with isobars $^{13}B$, $^{13}Be$ and $^{13}Li$ one have to expect synchronic one, two or more of **p→n** transitions with the following internal **n→p** "return" through the internal capture of $\pi^+$, $\mu^+$ or $e^+$.

According to *Writing I*, the charge lowering population of the $^{13}B$ states could look like this below:

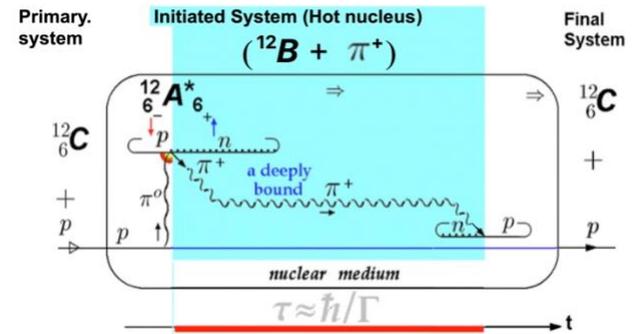

Fig.12.

According to *Writing II*, the charge lowering population of the $^{13}B$ states could look like this below:

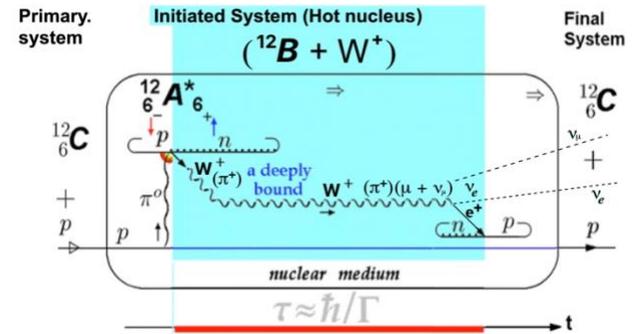

Fig.13.

One can show many similar Feynman diagrams, but more important is to find for sure what's behind of these EF-anomalies.

Looks like the issue of 80% of unrecognized structures was unclear until the CIRs concept was proposed and used in attempt to decode belonging of this massive pile of unknown peaks.

As is clear from Fig. 4, the CIRs concept result looks very promising, but where are the proofs that CIRs are real?

In confirmation of CIRs, the easiest way is to analyze the older precise data for presence of CIRs by the *Writing I* and *Writing II*.

## 5. The CIRs confirmations.

For the first let's start with the same $^{12}C$ nucleus but measured at Van-de-Graaff [21] in the energy range 4.7 MeV÷11.8 MeV under 27.6°, with $E_p$ step 2.5 keV.

The **Reading 0** (1) has to work first and show all resonances, corresponding to the real energy accessible products (nuclei and particles) on Fig. 14.

# Reading 0.

$$E_{pRF} = E^*_{LNP} + E^*_{RC} + \Delta M_{IS} - \Delta M_{PS}$$

proton c.m. energy

$E^*_{LNP}$ level energy of nucleus-product

$\Delta M_{IS}$ mass excess of initiated system

$\Delta M_{PS}$ mass excess of primary system

$\Delta M_{IS} = \Delta M_{TN} + \Delta m_{RC}$     $\Delta M_{PS} = \Delta M_{TN} + \Delta m_{BP}$

| SBC | Swint J.B., Barnard A.C.L., Clegg T.B. and J.L.Weil   NP 86, (1966), p.119–129 |

$\Delta M_{TN}$ — mass excess of target-nucleus
$\Delta M_{NP}$ — mass excess of nucleus-product
$\Delta m_{RC}$ — mass excess of residual product-cluster(s) or/and particle(s)
$\Delta m_{BP}$ — mass excess of beam-particle

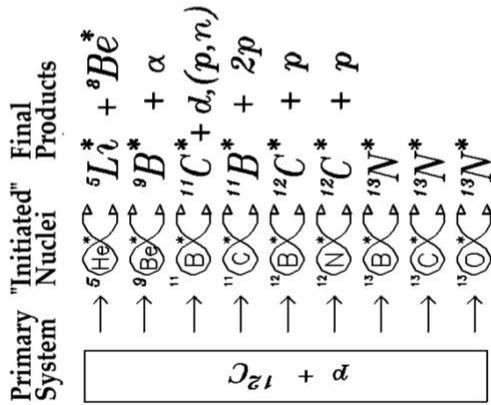

Primary System / "Initiated" Nuclei / Final Products:

- $^5\text{He} \rightarrow ^5\text{Li}^* + ^8\text{Be}^*$
- $^9\text{Be} \rightarrow ^9\text{B}^* + \alpha$
- $^{11}\text{B} \rightarrow ^{11}\text{C}^* + d, (p,n)$
- $^{11}\text{C} \rightarrow ^{11}\text{B}^* + 2p$
- $^{12}\text{B} \rightarrow ^{12}\text{C}^* + p$
- $^{12}\text{N} \rightarrow ^{12}\text{C}^* + p$
- $^{13}\text{B} \rightarrow ^{13}\text{N}^* + p$
- $^{13}\text{C} \rightarrow ^{13}\text{N}^*$
- $^{13}\text{O} \rightarrow ^{13}\text{N}^*$

$p + ^{12}C$

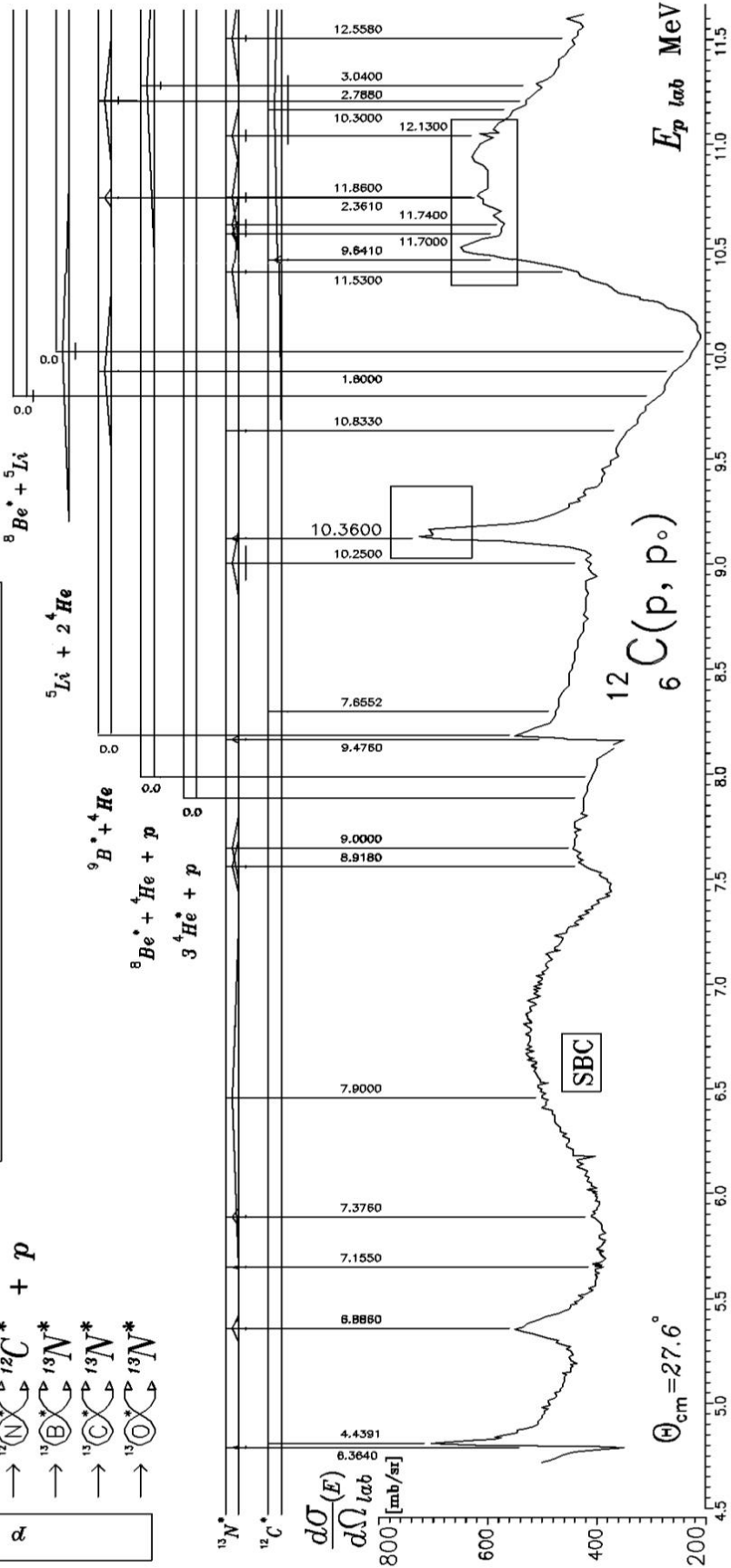

Fig. 14. Excitation function $^{12}C(p, p_\circ)$ measured in the energy range 4.7 MeV – 11.5 MeV with the energy step 2.5 keV. Population of states by Reading 0.

The precise excitation function from Van-de-Graaff measurement shows many resonances, filled by events, which are unsuccessful attempts to produce well-known nuclei and particles.

### $^{12}C$

Let's start with levels of $^{12}C$. First, it's a dazzling resonance of excited level 4.4391 MeV, then a modest peak of 7.6552 MeV, next one 9.6410 MeV and a wide one 10.3000 MeV.

### $^{13}N$

Next resonances correspond to (29) population and decay of $^{13}N$ states. First comes the bright cusp at 6.364 MeV, the very bright 6.8860 MeV, also several of little displayed 7.1550 MeV, 7.3760 MeV, 7.9000 MeV, 8.9180 MeV, 9.0000 MeV. Then again, a bright cusp at 9.4760 MeV. Next two very modest at 7.6552 MeV and at 10.2500 MeV. Then a dazzling one comes at 10.3600 MeV. Next one is almost not pronounced at 10.8330 MeV. Then it's hill made of several levels 11.5300 MeV, 11.7000 MeV, 11.7400 MeV, 11.8600 MeV, 12.1300 MeV, and 12.5580 MeV. This hill has some other anomalies – candidates to CIRs.

### $3\ ^4He + p$

This threshold is almost unpronounced at $E_p$=7.577 MeV. No levels excited.

### $^8Be + {}^4He + p$

The threshold is unpronounced $E_p$=7.95 MeV. One level gets populated with 3.0400 MeV

### $^9Be + {}^4He$

This threshold is a very bright one at $E_p$=8.35 MeV. First $^9Be$ level 1.600 MeV-almost nothing, and two others are brighter 2.3610 MeV, 2.7880 MeV.

### $^8Be + {}^5Li$

Threshold is almost unnoticed at $E_p$=9.793 MeV.

### $^5Li\ +\ 2\ ^4He$

Threshold is almost unnoticed at $E_p$=10.009 MeV.

## 5.1. CIRs by *Writing I*

Now let's see what will happen when *Writing I* is applied (Fig. 15).

### initiated ($^{13}C$)

First 12 levels do almost nothing, here they are: 8.2 MeV, 8.8600 MeV, 9.4998 MeV, 9.8970 MeV, 10.4650 MeV, 10.7530 MeV, 10.8180 MeV, 10.9960 MeV, 11.080 MeV, 11.7480 MeV, 11.8480 MeV, 11.95 MeV.

Then it's a bright peak **12.1060** MeV and a teil with excited levels 12.13 MeV, 12.14 MeV, 12.1870 MeV and 12.4360 MeV.

The level 13.00 MeV go unnoticed. Them several levels make a hill with 13.2800 MeV, 13.4100 MeV. Next levels are well visible **13.5700** MeV, **13.7800** MeV.

Obvious CIRs-I from $^{13}C$ population are the dazzling **12.1060** MeV, and also **13.5700** MeV and **13.7800** MeV.

## 5.2. CIRs by *Writing II*

The Fig. 16 shows population of initiated nuclei (g.s. or/and excited* levels) with contribution of the weak and electromagnetic forces.

### initiated ($^{13}C$)

First 14 levels of this initiated nucleus look not showing any noticeable anomalies in the EF. Here they are: 6.8640 MeV, 7.4920 MeV, 7.5470 MeV, 7.6860 MeV, 8.2000 MeV, 8.8600 MeV, 9.4998 MeV, 9.8970 MeV, 10.4650 MeV, 10.7530 MeV, 10.8180 MeV, 10.9960 MeV.

Next level forms a dazzling peak of level **11.080** MeV. The peak isn't sharp like others –the neutrino widening?

The following four levels again do not show any significant anomalies: 11.7480 MeV, 11.8480 MeV, 11.9500 MeV, 12.1060 MeV.

### initiated ($^9Be$) + $^4He$

This threshold is on the high peak tail and it doesn't show any anomaly. But there is one level with **1.6850** MeV.

### Initiated ($^5He$) + $^8Be$

This threshold shows a low peak at $E_p$=10.35 MeV.

### initiated ($^5He$) + $2\ ^4He$

This threshold shows a good peak at $E_p$=**10.825** MeV.

Obvious CIRs-II (with weak/electromagnetic interaction involved) from $^{13}C$ population are the dazzling **11.080** MeV, and also the ($^5He$) + $2\ ^4He$ threshold at $E_p$=**10.825** MeV.

## 5.3. The CIRs in the present EF: **6 peaks**

Three peaks from *Writing I* (Fig. 15): **12.1060** MeV, **13.5700** MeV and also **13.7800** MeV.

Three peaks from *Writing II* (Fig. 16): **11.080** MeV, **1.6850** MeV and one at $E_p$=**10.825** MeV.

# Writing I.

$$E_{pRF} = E^*_{LNP} + E^*_{RC} + \Delta M_{IS} - \Delta M_{PS}$$

proton c.m. energy / level energy of nucleus-product / mass excess of initiated system / mass excess of primary system

$\Delta M_{IS} = \Delta M_{NP} + \Delta m_{RC}$     $\Delta M_{PS} = \Delta M_{TN} + \Delta m_{BP}$

$\Delta M_{NP}$ - mass excess of nucleus-product  
$\Delta m_{RC}$ - mass excess of residual product-cluster(s) or/and particle(s)  
$\Delta M_{TN}$ - mass excess of target-nucleus  
$\Delta m_{BP}$ - mass excess of beam-particle

| SBC | Swint J.B., Barnard A.C.L., Clegg T.B. and J.L.Weil NP 86, (1966), p.119-129 |

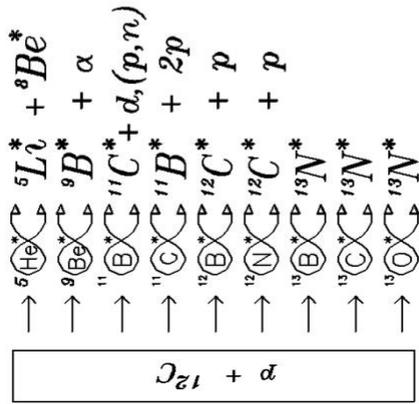

Primary "Initiated" Final  
System     Nuclei    Products

$p + {}^{12}C$ →
- ${}^{5}He^* \to {}^{5}Li^* + {}^{8}Be^*$
- ${}^{9}Be^* \to {}^{9}B^* + \alpha$
- ${}^{11}B^* \to {}^{11}C^* + d, (p,n)$
- ${}^{11}C^* \to {}^{11}B^* + 2p$
- ${}^{12}B^* \to {}^{12}C^* + p$
- ${}^{12}N^* \to {}^{12}C^* + p$
- ${}^{13}B^* \to {}^{13}N^* $
- ${}^{13}C^* \to {}^{13}N^*$
- ${}^{13}O^* \to {}^{13}N^*$

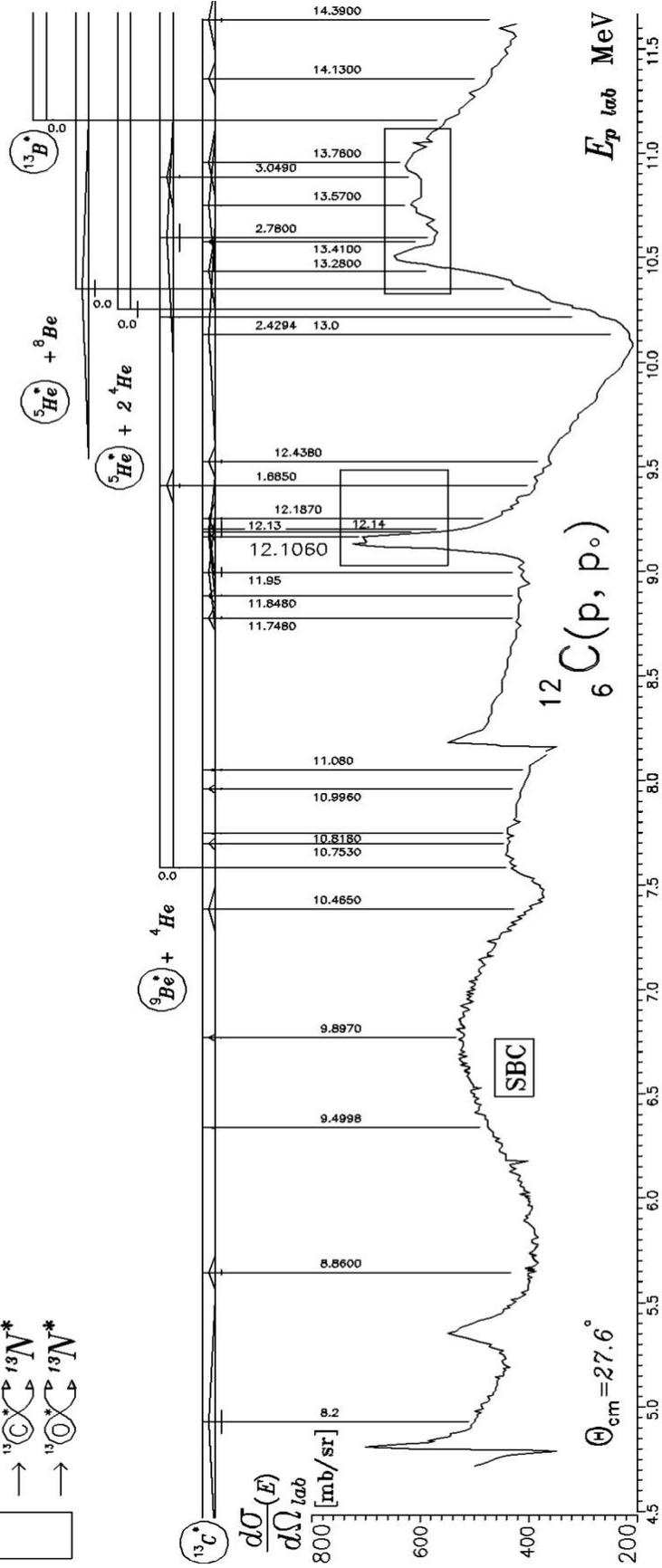

Fig. 15. Excitation function of ${}^{12}C(p, p_0)$ with CIR-resonances according to the Writing I (only strong forces involved).

# Writing II.

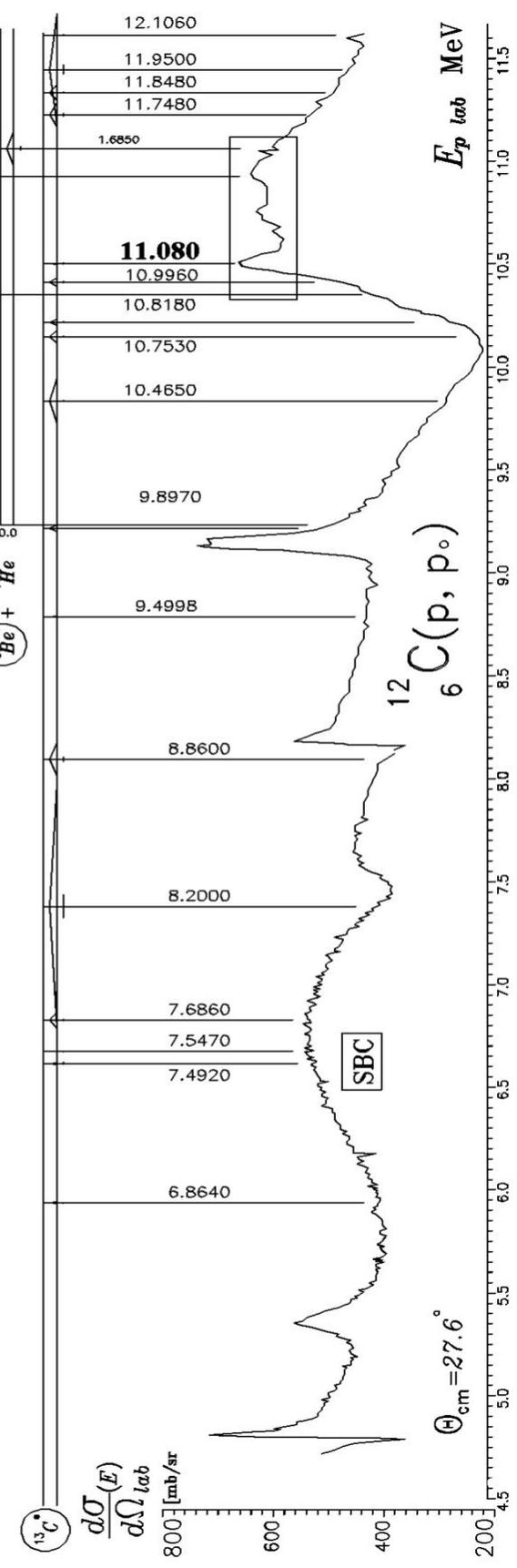

$$E_{p,RF} = E^*_{LIN} + E^*_{RC} + \Delta M_{FS} - \Delta M_{PS} + E_{NUC}$$

| | |
|---|---|
| $E_{p,RF}$ — proton c.m. energy | $E^*_{LIN}$ — level energy of initiated nucleus & residual cluster(s) |
| $\Delta M_{FS}$ — mass excess of final system | $\Delta M_{PS}$ — mass excess of primary system |
| | $E_{NUC}$ — nuclear energy |

$\Delta M_{FS} = \Delta M_{FN} + \Delta m_{PP}$   $\Delta M_{PS} = \Delta M_{TN} + \Delta m_{BP}$

$\Delta M_{FN}$ — mass defect of final nucleus   $\Delta M_{TN}$ — mass defect of target-nucleus

$\Delta m_{PP}$ — mass defect of particle-product   $\Delta m_{BP}$ — mass defect of beam particle

| SBC | Swint J.B., Barnard A.C.L., Clegg T.B. and J.L.Weil  NP 86, (1966), p.119–129 |
|---|---|

I, F — the initiated and final nuclei-products respect.

$E_{NUC} = \Delta E_{IN} - \Delta E_{FN}$

$\Delta M_{IN}$ — mass defect of initiated nucleus

$\Delta E_{IN} = \Delta M_{IN} - Z_I \Delta m_p - N_I \Delta m_n - 0.6 * Z_I * (Z_I - 1)/\sqrt[3]{A_I}$

$\Delta E_{FN} = \Delta M_{FN} - Z_F \Delta m_p - N_F \Delta m_n - 0.6 * Z_F * (Z_F - 1)/\sqrt[3]{A_F}$

Z, N — numb. n, p, $\Delta m_p$, $\Delta m_n$ — their mass defects

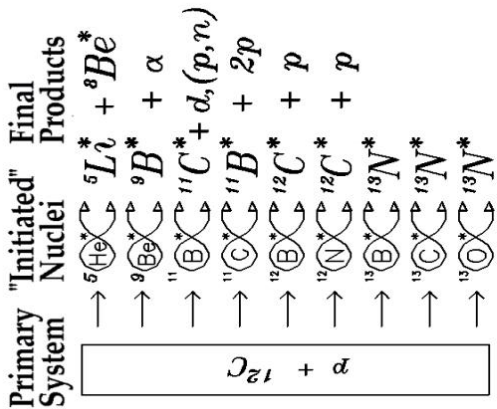

| Primary System | "Initiated" Nuclei | Final Products |
|---|---|---|
| $p + {}^{12}C$ | ${}^5He^*$ | ${}^5Li^* + {}^8Be$ |
| | ${}^9Be^*$ | ${}^9B^* + \alpha$ |
| | ${}^{11}B^*$ | ${}^{11}C^* + d, (p,n)$ |
| | ${}^{11}C^*$ | ${}^{11}B^* + 2p$ |
| | ${}^{12}B^*$ | ${}^{12}C^* + p$ |
| | ${}^{12}N^*$ | ${}^{12}C^* + p$ |
| | ${}^{13}B^*$ | ${}^{13}N^*$ |
| | ${}^{13}C^*$ | ${}^{13}N^*$ |
| | ${}^{13}O^*$ | ${}^{13}N^*$ |

Fig. 16. Excitation function of ${}^{12}C(p, p_0)$ with CIR-resonances according to the Writing II (with weak and electromagnetic forces involved).

## 6. The CIRs in the EF of $^{12}C(d, d_o)$.

Let's search for presence of the CIRs in the precise EF measured for $^{12}C(d, d_o)$ for the deuteron energy range 0.9 MeV ÷ 5 MeV [22].
The Fig. 18 shows results of the Reading 0 for the compound nucleus $^{14}N$:

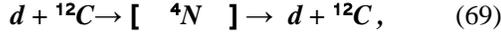
$$d + {}^{12}C \to [\quad {}^4N \quad] \to d + {}^{12}C, \qquad (69)$$

Almost all anomalies are in a precise agreement with verticals corresponding to the excited levels of $^{14}N$.
**"Unnoticed" are only 10 anomalies: at 0.95 MeV, 1.28 MeV, 1.84 MeV, 2.22 MeV, 2.32 MeV, 2.45 MeV, 2.625 MeV, 2.75 MeV, 4.15 MeV and 4.76 MeV.**

### 6.1. CIRs by *Writing I*

The Fig. 19 (the lower side) shows the found CIRs, according to the *Writing I*, in this energy range.

There are some lighter combinations:
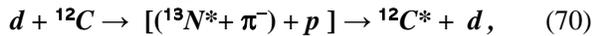
$$d + {}^{12}C \to [({}^{13}N^* + \pi^-) + p] \to {}^{12}C^* + d, \qquad (70)$$
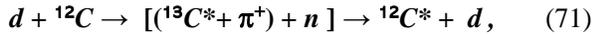
$$d + {}^{12}C \to [({}^{13}C^* + \pi^+) + n] \to {}^{12}C^* + d, \qquad (71)$$

Also the (69) splits in two branches:

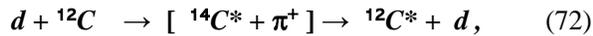
$$d + {}^{12}C \to [{}^{14}C^* + \pi^+] \to {}^{12}C^* + d, \qquad (72)$$
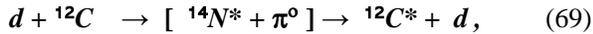
$$d + {}^{12}C \to [{}^{14}N^* + \pi^0] \to {}^{12}C^* + d, \qquad (69)$$
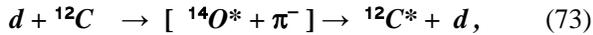
$$d + {}^{12}C \to [{}^{14}O^* + \pi^-] \to {}^{12}C^* + d, \qquad (73)$$

On the Fig. 17 is shown an example of the Feynman diagram for reaction (72) with initiated ($^{14}C + \pi^+$):

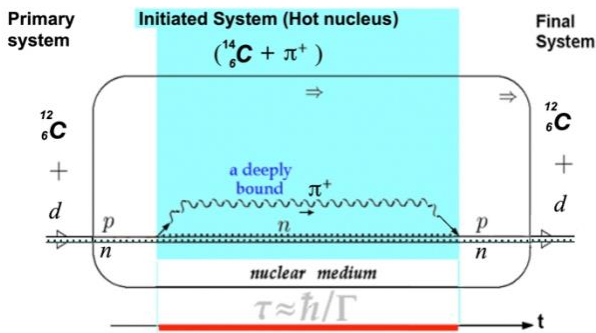
Fig.17.

($^{13}C$)
There is a noticeable immediate precise agreement for $^{13}C$ (71) of level 3.6845 MeV at $E_d$ =2.625 MeV.

($^{13}N$)
Next one is for $^{13}N^*$ (70) of level 3.547 MeV at $E_d$ = 2.817 MeV.

($^{14}O$)
Then for $^{14}O^*$ (73) level 7.768 at brightest peak $E_d$ = 2.505 MeV, it's a perfect match! Also, a noticeable one is the level 9.715 MeV at $E_d$ =4.76 MeV.

($^{14}C$)
Really, a dazzling one is $^{14}C^*$ (72) with levels 10.4250 MeV at $E_d$ =0.95 MeV, level 11.5000 MeV at $E_d$ =2.22 MeV, and level 13.7000 MeV at $E_d$ =4.76 MeV.

### 6.2. CIRs by *Writing II*

The found CIRs by *Writing II,* for the weak and electromagnetic forces involved, are very bright too (Fig.19, the upper side).

($^{14}O$)
The $^{14}O^*$ level 9.915 is dazzling at brightest peak on $E_d$ =4.76 MeV. Then it's level 10.890 MeV at $E_d$ =2.97 MeV. What a match of level 11.240 at $E_d$ =3.385 MeV!.

($^{14}C$)
In a perfect match is $^{14}C^*$ level 9.7460 MeV at $E_d$ =2.75 MeV, then also level 9.8010 MeV at $E_d$ =2.82 MeV.

### 6.3. The CIRs in $^{12}C(d, d_o)$ EF: **11 peaks**

Six peaks from *Writing I* (Fig. 19, the lower section) at $E_d$ =0.95 MeV; 2.22 MeV; 2.625 MeV; 2.817 MeV; 2.505 MeV; 4.76 MeV.
Five peaks from *Writing II* (Fig. 19, the upper section) at $E_d$ =2.75 MeV; 2.82 MeV; 2.97 MeV; 3.385 MeV; 4.76 MeV.

Overall, **eleven CIRs** in the studied $^{12}C(d, d_o)$ excitation function [22] for deuteron energy range 0.9 MeV÷5 MeV.

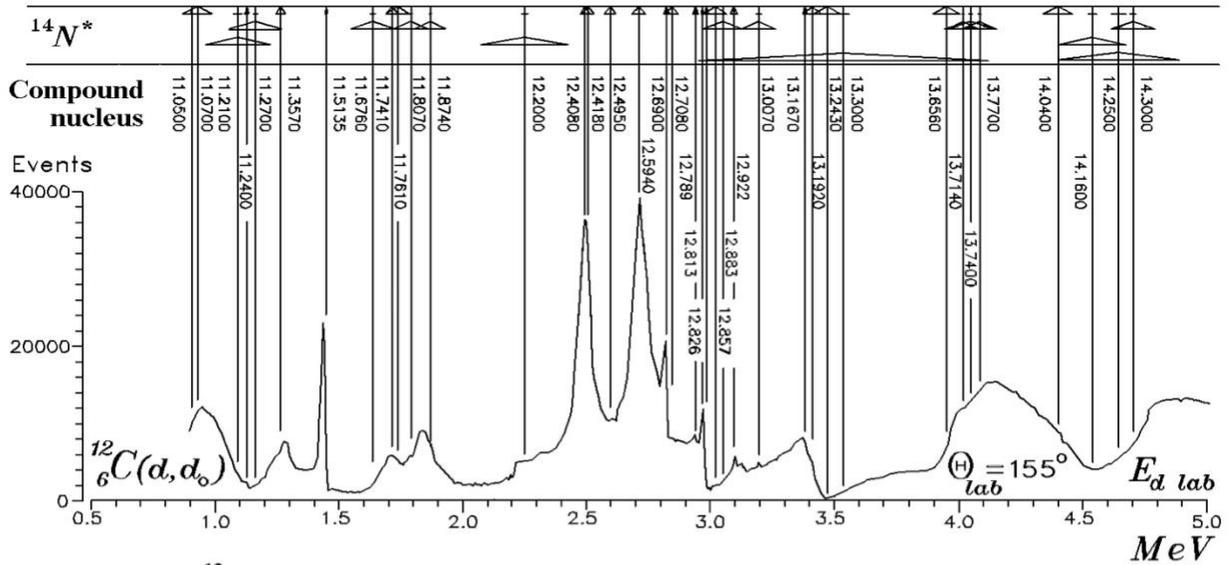

Fig. 18. The EF $^{12}C(d, d_o)$ for the deuteron energy range 0.9 MeV - 5 MeV according to **Reading 0**, (published at [27] Jeronymo et al., Nucl. Phys. 43 (1963) p. 417-423)

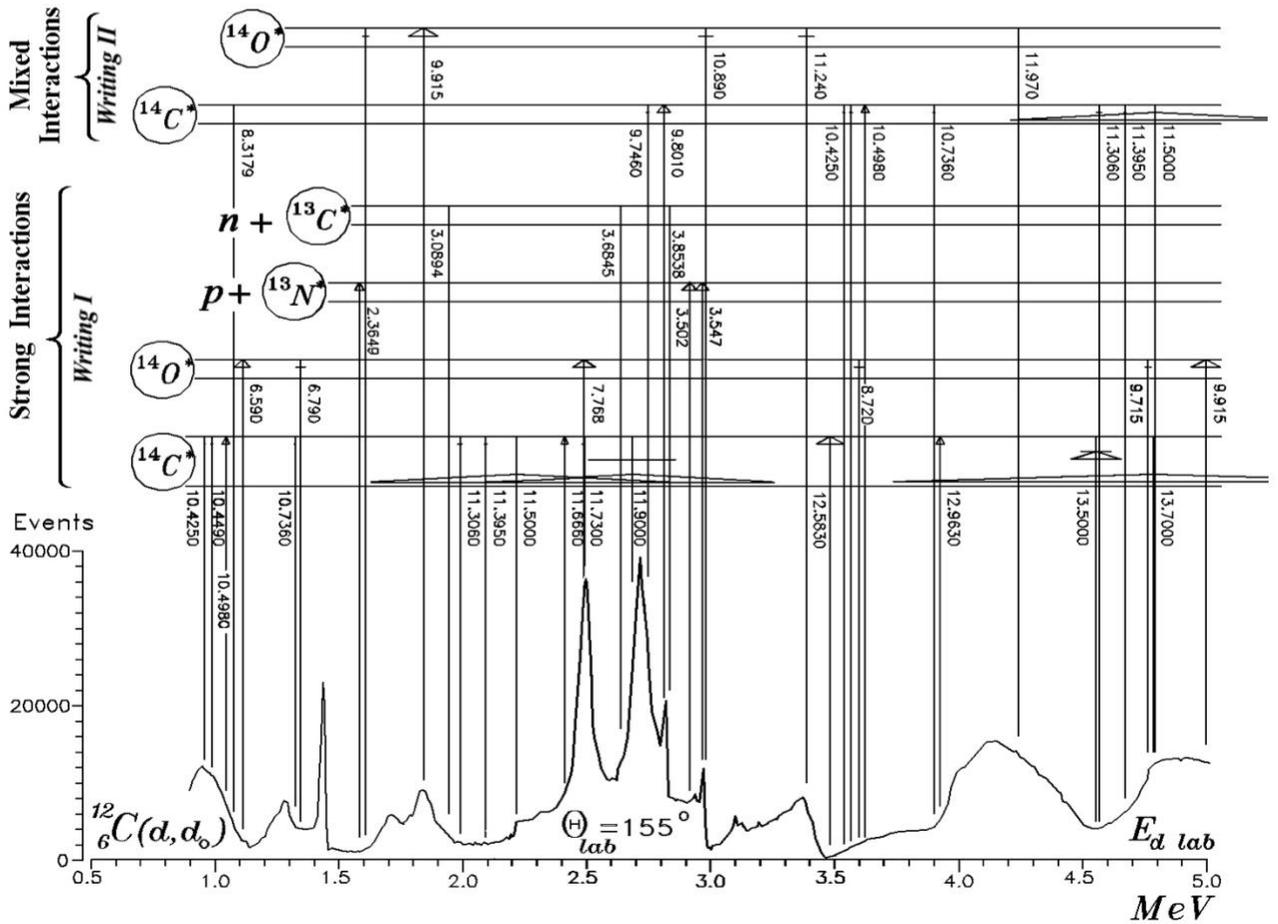

Fig. 19. The EF $^{12}C(d, d_o)$ for the deuteron energy range 0.9 MeV - 5 MeV according to *Writing I* (Strong) and *Writing II* (Mixed)  (published at [27] Jeronymo et al., Nucl. Phys. 43 (1963) p. 417-423)

## 7. The precise EF of $^7Li(p, p_o)$

Let's first search for final products traces in this precise EF measured [23] for $^7Li(p, p_o)$ for the proton energy range 1.2 MeV ÷ 3.1 MeV.
Results of the **Reading 0** for the final products are shown below and on Fig. 21:

$$p + ^7Li \rightarrow [\ ^8Be\ ] \rightarrow p + ^7Li, \qquad (74)$$
$$p + ^7Li \rightarrow [\ ^7Be + n\ ] \rightarrow p + ^7Li, \qquad (75)$$
$$p + ^7Li \rightarrow [\ ^4He + p + t\ ] \rightarrow p + ^7Li, \qquad (76)$$

The vertical pointers precisely fit the displayed anomalies in the EF of the $p + ^7Li$ elastic scattering.
Especially noticeable are resonances that correspond to (74) population of the $^8Be$ nucleus in excited states: 18.91 MeV, 19.07 MeV and 19.24 MeV.
Also, population (75) of the combination $[n + ^7Be\ g.s/*]$ is noticeable because of the g.s and first $^7Be$ excited state with energy 0.42908 MeV.
The g.s of $^4He$ in combination (76) $[^4He + p + t]$ got almost unnoticed.
The only obviously unknown peak is at $E_p$=2.02 MeV.

### 7.1. CIRs in $^7Li(p, p_o)$ by *Writing I*

Let's search, in this EF, for presence of the CIRs, that correspond to strong forces (*Writing I*).

The right side of Fig. 21 shows the Combinative Isobaric Resonance's contribution to the EF-shape.

The expected population of the initiated systems are below:

$$p + ^7Li \rightarrow [\ ^7Li\ g.s\ or\ * + (\pi^+ + n)\ ] \rightarrow p + ^7Li, \quad (77)$$
$$p + ^7Li \rightarrow [\ (^8Li\ g.s\ or\ * + \pi^+)\ ] \rightarrow p + ^7Li, \quad (78)$$
$$p + ^7Li \rightarrow [\ (^8B\ g.s\ or\ * + \pi^-)\ ] \rightarrow p + ^7Li, \quad (79)$$

Immediate precise match shows the (77) $[(^7Li^*) + n]$ combination in the first ($^7Li^*$) excited state with energy 0.4776 MeV.

The single bright CIR at $E_p$=2.02 MeV corresponds to (77) population of first level of $^7Li^*$ with $p \rightarrow (\pi^+ + n)$ transition in the nuclear medium (Fig.20, below):

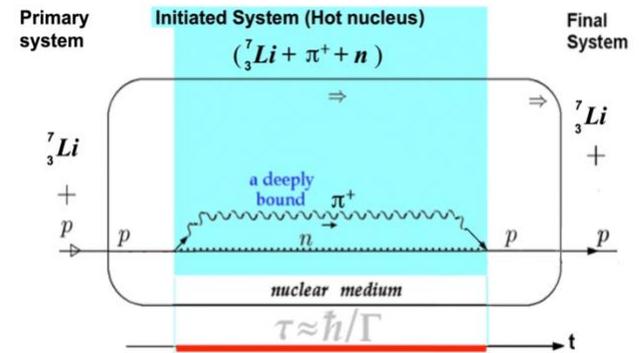

Fig. 20. Feynman diagram for the $[^7Li^* + (n + \pi^+)]$ -IS.

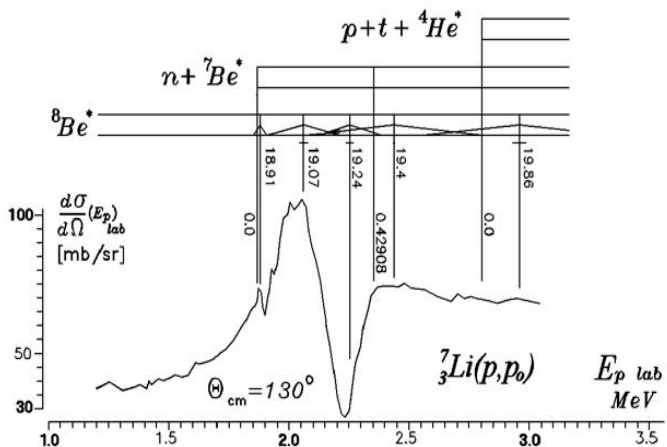
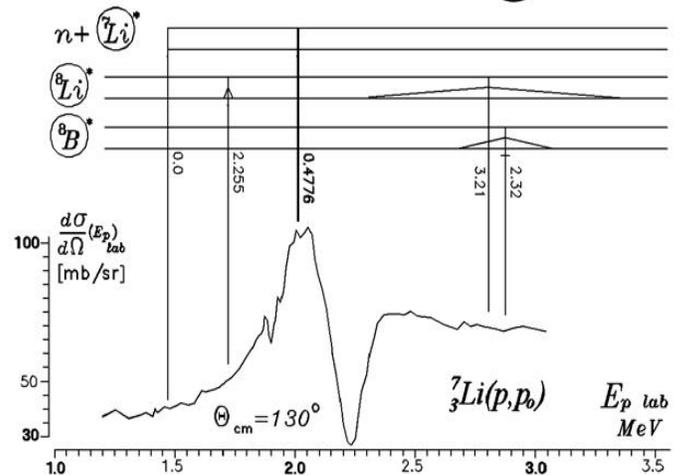

Fig. 21. Excitation Function of $^7Li(p, p_o)$ by **Reading 0** and also analyzed by **Writing I** ([28] Phys. Rev. 101.114 (1956))

# 8. Conclusion

For over 50 years of nuclear physics evolution, all it was heard for many times: Isobar-analog states and resonances, the "doorway states", the isospin excitations and mixing…

The Combinative Isobaric Resonances, the CIRs concept, include all of them and briefly explains
- why CIRs exist and what they are,
- where and how CIRs appear and look, and
- what are the obvious consequences of the CIRs.

**By studying the CIRs one can get very reach and important information on the intranuclear dynamics in details with all involved things!**
All this again says, the NATURE is much richer than one could imagine.

Next preprint would be devoted to the non-obvious consequences of the CIRs and where first to look for more important revisions and news.

# 9. Acknowledgements

I would like to express my big thanks to Dr. of Sciences Avas V. Khugaev (INP, SA, RUz, Ulugbek), Dr. of Sciences Avazvek K. Nasirov (BLTP, JINR), and also Dr. of Sciences Gurgen G. Adamian (BLTP, JINR).

A special big thanks to professor Dr. Fangil Gareev (LNR, JINR, Russia), who discussed the CIRs with me. Big thanks to Dr. Gennady A. Radiyk (INP, Tashkent) who many times discussed the MSS and its results with me.

My sincere gratitude to BU-scientists professor James P. Miller and professor Robert M. Carey.

# References.